\documentclass[twocolumn,showpacs,superscriptaddress,amsmath,amssymb,prl,10pt,aps,longbibliography]{revtex4-1}
\usepackage{graphicx}
\usepackage{dcolumn}
\usepackage{bm}
\usepackage{upgreek}

\usepackage{subfigure}
\usepackage[utf8]{inputenc}
\usepackage{multirow}
\usepackage{soul}
\usepackage{float}
\usepackage[normalem]{ulem}
\usepackage[usenames,dvipsnames]{color}
\usepackage[export]{adjustbox}
\usepackage{hyperref}

\definecolor{tangerine}{rgb}{0.944,0.522,0}

\def\Zstar{\boldsymbol{\mathsf{Z}}^*}

\def\barR{\overline{\mathbf R}}
\def\barC{\overline{\mathcal{C}}}

\newcommand{\editor}[2]{%
  \expandafter\newcommand\csname #1note\endcsname[1]{%
    \textcolor{#2}{(\textbf{#1:} ##1)}}%
  \expandafter\newcommand\csname #1\endcsname[1]{%
    \textcolor{#2}{##1}}%
  \expandafter\newcommand\csname #1cancel\endcsname[1]{%
    \textcolor{#2}{\sout{##1}}}%
  \expandafter\newcommand\csname #1change\endcsname[2]{%
    \textcolor{#2}{\sout{##1} ##2}}%
  \newenvironment{#1text}{\color{#2}}{\color{black}}
}

\editor{FG}{blue}
\editor{SB}{tangerine}
\editor{RS}{cyan}

\begin{document}

\title {Topological quantisation and gauge invariance \\of charge transport in liquid insulators}
\author{Federico Grasselli}
\affiliation{SISSA -- Scuola Internazionale Superiore di Studi Avanzati, Via Bonomea 265, 34136 Trieste, Italy}
\affiliation{\textsc{MaX} EU Centre of Excellence for Supercomputing Applications @SISSA}
\author{Stefano Baroni}
\email{baroni@sissa.it}
\affiliation{SISSA -- Scuola Internazionale Superiore di Studi Avanzati, Via Bonomea 265, 34136 Trieste, Italy}
\affiliation{\textsc{MaX} EU Centre of Excellence for Supercomputing Applications @SISSA}
\affiliation{CNR -- IOM DEMOCRITOS @SISSA}
\date{\today}

\begin{abstract}
According to the Green-Kubo theory of linear response, the conductivity of an electronically gapped liquid can be expressed in terms of the time correlations of the adiabatic charge flux, which is determined by the atomic velocities and Born effective charges. We show that topological quantisation of adiabatic charge transport and gauge invariance of transport coefficients allow one to rigorously express the electrical conductivity of an insulating fluid in terms of integer-valued, scalar, and time-independent atomic oxidation numbers, instead of real-valued, tensor, and time-dependent Born charges.
\end{abstract}


\maketitle

Electronically insulating liquids can carry an electric current in response to an applied electric field, as their atomic or molecular constituents may carry a charge. Common examples are ionic solutions, molten salts, and ionic liquids. Within the Green-Kubo (GK) theory of linear response \cite{Green1952,*Green1954,Kubo1957a,*Kubo1957b}, the electrical conductivity of a classical fluid is given by the celebrated GK formula:
\begin{equation}
\sigma =\frac{{\Omega}}{3k_BT} \int_0^\infty \langle \mathbf J (t) \cdot \mathbf J (0) \rangle \, dt, \label{eq:GK}
\end{equation}
where $\Omega$ and $T$ are the system volume and temperature, $k_B$ is the Boltzmann constant, $\langle\cdot\rangle$ indicates an equilibrium ensemble {average}, and $ \mathbf J(t) = {\frac{1}{\Omega}} \sum_i q_i \mathbf v_i(t) $ is the electric charge flux, $\mathbf v_i$ and  $q_i$ being the velocity and classical charge of the $i$-th atom, and the sum extends over the $N$ atoms of the system.

The situation is not nearly as clear when a quantum mechanical picture of the interatomic forces is adopted, because atomic charges are ill defined in this case. It seems therefore that the adoption of any of the many available and inevitably arbitrary definitions of atomic charge would lead to a different expression for the electric charge flux and value for the conductivity. This ambiguity is lifted by considering that the charge flux is the time derivative of the macroscopic polarisation, $\mathbf P${: $\mathbf J=\dot{\mathbf P}$}. In the adiabatic approximation, $\mathbf{P}$ depends on time only through the nuclear coordinates, so that its time derivative reads:
\begin{equation}
	\mathbf J(t) = {\frac{1}{\Omega}} \sum_i \Zstar_i(t) \cdot \mathbf v_i(t), \label{eq:current}
\end{equation}
where the Born effective charge, $\Zstar_i$, is a tensor whose components are derivatives of the system's dipole{, $\bm{\upmu} = \Omega \mathbf{P}$,}  with respect to atomic displacements: $Z^*_{i\alpha\beta} = \frac{\partial {\mu}_\alpha}{\partial r_{i\beta}}$, $\mathbf r_i$ being the position of the $i$-th atom. The implementation of the Kubo formula, Eq.~\eqref{eq:GK}, from first principles thus requires the numerical evaluation of the Born effective charges along a molecular trajectory, using either a linear-response \cite{Baroni2001} or a Berry-phase \cite{Resta2010,Vanderbilt2018} approach.

This procedure was implemented, \emph{e.g.}, in Ref. \onlinecite{French2011} in the case of partially ionic water, a state occurring at the high-PT conditions to be found in the icy giants' interior. In that paper an outstanding conundrum was identified, in that \emph{interestingly, the use of predefined constant charges can yield the same conductivity as is found with the fully time-dependent charge tensors} (verbatim). Even more interestingly, those \emph{predefined constant charges} coincide with what chemical intuition would suggest for the oxidation numbers of O ($q_\mathrm{O}=-2$) and H ($q_\mathrm{H}=+1$). We note that a similar poser occurs in the electrical properties of atomically neutral fluids: how is it that a vanishing conductivity can result through the GK formula, Eq.~\eqref{eq:GK}, from the time series of a non-vanishing charge flux, Eq.~\eqref{eq:current}? The question, then, naturally arises: are these numerical coincidences, or the consequence of a deep, hitherto unrecognised, invariance principle? In the latter case, does this principle stem from a fundamental theory or from just an approximation of some sort? Nearly at the same time, another paper \cite{Jiang2012} appeared where, based on the modern theory of polarisation \cite{Resta2010,Vanderbilt2018}, it was shown that oxidation states can be rigorously associated with individual atoms in insulating crystals, such that the total charge transported by the displacement of an atomic sublattice by a lattice vector is an integer. While this finding certainly bears some relevance to the solution of our conundrum, the extent to which it can be generalised to liquids and the impact it can have on transport properties is not evident at all.

In this work we demonstrate that the above coincidences are by no means such, but they rather stem from the topological properties of the electronic structure of insulating materials. To this end, we first derive a rigorous definition of atomic oxidation numbers in \textit{liquid} insulators, based on purely topological arguments, and discuss their general properties, such as quantisation and additivity. We then show that these two concepts can be combined with a recently discovered \emph{gauge invariance} of transport coefficients \cite{Marcolongo2016,Ercole2016,Baroni2018} in such a way that defining the charge flux, Eq.~\eqref{eq:current}, in terms of these integer, constant, and scalar numbers, instead of real, time-dependent, and tensor Born effective charges, results in the same conductivity, as computed from Eq.~\eqref{eq:GK}, thus solving the conundrum highlighted in Ref. \onlinecite{French2011}. Our theoretical results are demonstrated numerically {on a model of molten potassium chloride (KCl). We computed} the charge transported along a number of representative periodic paths involving the net displacement of one or two atoms{, and compared} the electric conductivities {extracted} from \emph{ab initio} (AI) molecular dynamics (MD) and the GK formula, {employing} alternatively the Born effective charges and the newly defined atomic oxidation numbers.

\begin{figure}
    \centering
    \includegraphics[width = 0.9\linewidth]{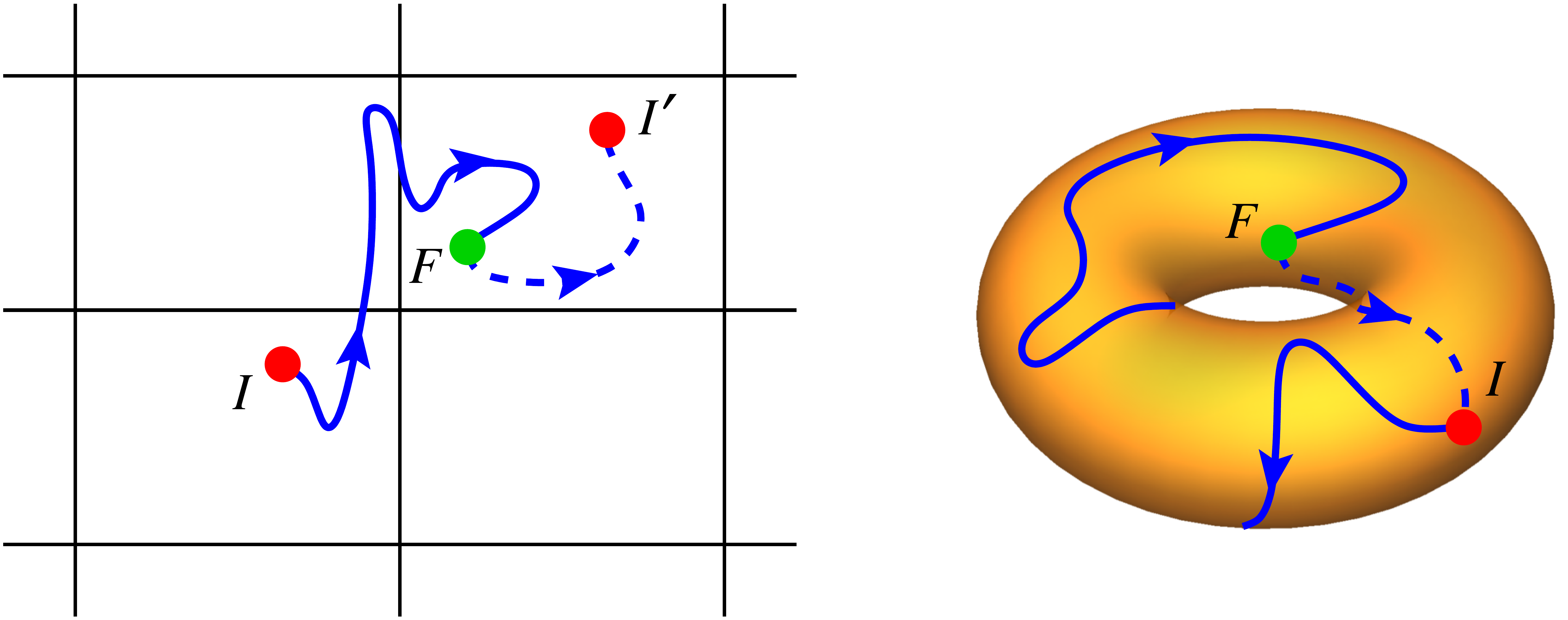}
    \caption{Paths in the periodic nuclear configuration space. A molecular trajectory, $\mathcal{C}_{IF}$, originating at point $I$ of the $3N$-dimensional unit cell of a periodic lattice and ending at point $F$ of a different cell (left) is topologically equivalent to an open path path on a torus (right). $\mathcal{C}_{IF}$ can be seen as the concatenation of a path joining $I$ with its periodic image, $I'$, lying in the same cell as $F$, with a path joining $I'$ with $F$: $\mathcal{C}_{IF}=\mathcal{C}_{II'}\circ\mathcal{C}_{I'F}$. The image of $\mathcal{C}_{II'}$ on the torus is a loop (right). In this case, the winding numbers of the loop are $\mathbf{n}=(1,1)$, whereas the distance between $I'$ and $F$, $R_{I'F}=|\mathbf{R}_{F}-\mathbf{R}_{I'}|$ is bounded by the size of the unit cell.}
    \label{fig:ciambella}
\end{figure}

\section{Theory}
For the purposes of our work, it is expedient to express the conductivity through the Einstein-Helfand relation \cite{Helfand1960}, in terms of the slope of the mean square \emph{dipole displacement}, {$\Delta\bm{\upmu}(t)$}, as a function of time \cite{Baroni2018}:

\begin{gather}
   \sigma = \frac{1}{3 \Omega k_B T} \lim_{t \to \infty} \frac{\left \langle |\Delta{\bm{\upmu}}(t)|^2 \right\rangle}{2 t }, \label{eq:EH} \\
   {\Delta \bm{\upmu} = \Omega \int_0^t \mathbf{J}(t')dt'.} \label{eq:DeltaP}
\end{gather}
It is easily seen that any two expressions of the dipole displacement that differ by a bounded vector result in the same value of the electric conductivity, according to Eq.~\eqref{eq:EH}. This important property lies at the heart of the recently discovered \emph{gauge invariance} of transport coefficients \cite{Marcolongo2014,Ercole2016}. {In a nutshell, by gauge invariance we mean that transport coefficients are largely independent of the detailed form of the local representation of the conserved quantity (mass, energy, charge) being transported. This may apply to both continuous representations (densities) or to discrete ones in terms of atomic partitions. In the present case, we show that the total electronic charge of a system can be partitioned into suitably defined constant}
integer atomic oxidation numbers, $\{q_i\}$,
such that the dipole displacement computed from them, $\Delta {\bm{\upmu}}'(t)=\sum_i q_i \int_0^t  \mathbf{v}_i(t')dt'$, differs from $\Delta {\bm{\upmu}}(t)$ by a bounded vector, thus resulting in the same electric conductivity, according to Eq.~\eqref{eq:EH}.

When simulating dynamical phenomena in liquids, the system size has to be larger not only than the relevant correlation lengths, but also than the various diffusion lengths, \emph{i.e.} the distances travelled by each atom before it looses memory of its own velocity. When these requirements are met, equilibrium properties are independent of the boundary conditions adopted for the numerical simulation, and periodic boundary conditions (PBC) are normally chosen, because they minimise finite-size effects. When evaluating Eq.~\eqref{eq:EH} from MD simulations, the use of PBC is not only a matter of practical convenience, but also one of principle. In fact, when using reflecting or open boundary conditions, the $t\to\infty$ limit in Eq.~\eqref{eq:EH} vanishes for any finite system size, and it obviously does not commute with the thermodynamic limit. Using PBC and the definition \eqref{eq:DeltaP} for the dipole displacement, instead, the former limit is finite for any sample size and it commutes with the latter; PBC are thus the only ones able to sustain a steady-state charge flux \cite{RestaJuelich}. Our aim is to demonstrate that Eq.~\eqref{eq:EH} is unaffected if we replace the Born effective charges in the definition of the charge flux, Eq.~\eqref{eq:current}, with suitably defined atomic oxidation numbers. To achieve this goal, our reasoning will proceed using PBC, for which the long-time and large-size limits commute. Our conclusions, as well as the definition of oxidation numbers on which they stand, do not depend on the system size. We argue therefore that they hold in the thermodynamic limit as well.

A molecular trajectory with end points  $I\doteq\mathbf{R}(0)$ and  $F\doteq\mathbf{R}(\tau)$ is an open path, $\mathcal{C}_{IF}$, in the atomic configuration space (ACS), parametrised by time. While using PBC we will refer to $\mathcal{C}_{IF}$ as the path generated by the \emph{unwrapped} trajectory, $\mathbf{ R}(t)=\mathbf{R}(0) + \int_0^t \mathbf{V}(t') dt' \in \mathbb{R}^{3N}$,
where $\mathbf{V} = (\mathbf{v}_1, \ldots \mathbf{v}_N)$. 
The total dipole displaced along a trajectory, $\Delta{\bm{\upmu}}(\tau) = {\Omega} \int_0^\tau \mathbf{J}(t)dt=\int_I^F d{\bm{\upmu}} $ does not hinge on the time dependence of the trajectory, but only on the path and will be indicated with the shorthand $\Delta{\bm{\upmu}}_{IF}$. Because of this, if a trajectory is split into two paths,
$\mathcal{C}_{IF} = \mathcal{C}_{IX} \circ \mathcal{C}_{XF}$, the corresponding dipole results to be the sum of the dipoles associated to the two segments. In order to obtain the needed topological invariant, given a path corresponding to a physical trajectory, $\mathcal{C}_{IF}$, we first define a second path, $\mathcal{C}_{FI'}$, joining the final point of the molecular trajectory, $F$, with the periodic image of the initial point, $I'$, and lying entirely in the same unit cell, as illustrated in the left panel of Figure \ref{fig:ciambella}. Because of the above, evidently, one has:
$\Delta {\bm{\upmu}}_{IF} = \Delta {\bm{\upmu}}_{II'} + \Delta{\bm{\upmu}}_{I'F}$; furthermore,  $\Delta {\bm{\upmu}}_{I'F}$ is bounded. Therefore, in the long-time limit, the dipole displaced along $\mathcal{C}_{IF}$ and $\mathcal{C}_{II'}$ asymptotically coincide, and all we have to do is demonstrate that the latter can be expressed in terms of suitable integer topological invariants.

In order to achieve this goal, we consider the electronic Hamiltonian of the system, $\hat{H}(\lambda)$, as a function of a parameter $\lambda$, say $\in [0,1]$, labelling the atomic configuration along $\mathcal{C}_{II'}$. Evidently, $\hat{H}(\lambda)$ is periodic because the end points of the path are one a periodic image of the other: $\hat{H}(1)=\hat{H}(0)$. By making the assumption that the system's ground state stays gapped and non-degenerate all along $\mathcal{C}_{II'}$, Thouless' theorem on adiabatic charge transport \cite{Thouless1983,Pendry1984} ensures that the $\alpha$-th component of the total dipole displaced along $\mathcal{C}_{II'}$ is a multiple integer of the size, $\ell$, of the unit cell, which we assume to be cubic:
\begin{equation}
     Q_\alpha \doteq\frac{1}{\ell} \int_{\mathcal{C}_{II'}} { d \mu_\alpha }   \in  \mathbb Z. \label{eq:Thouless}
\end{equation}
{We stress that, strictly speaking, the quantisation condition expressed by Eq.~\eqref{eq:Thouless} only holds in the large-$\ell$ limit, when PBC are imposed to the electronic orbitals ($\Gamma$-point sampling) \cite{RESTA1998}. In practice, a system size of a few dozen atoms is large enough to guarantee that quantisation holds to two decimal digits.}
The $Q_\alpha$ charges are continuous functionals of the path that, being integer-valued, can only coincide for any two paths that can be continuously deformed into one another. We remark that in order for this to be possible, these deformations must be performed without {ever closing the electronic gap}. We now show that, under general assumptions, the $Q$ charges can be expressed as linear combinations of integer numbers---which will be identified with the atomic oxidation numbers---with integer coefficients.

A key element to accomplish the desired result is to consider the ACS from a topological point of view, whereby periodic boundary conditions make it topologically equivalent to the $3N$-dimensional torus $\mathbb{T}^{3N}$. It is thus convenient to map the path $\mathcal{C}_{II'}$ onto $\mathbb{T}^{3N}$, where the images of the end points $I$ and $I'$ coincide, so that its own image, $\overline{\mathcal{C}}_{I}$, is a closed path (\textit{loop}). Here and in the following we denote the images on $\mathbb{T}^{3N}$ of the unwrapped points and trajectories in $\mathbb{R}^{3N}$ by an overline, as in $\barR$ and $\barC$.
Loops are a standard tool to classify a topological space: this can be in fact characterised by its \emph{fundamental group}, defined as the set of homotopy classes of loops containing $\barR(0)$ as \emph{base point}, and equipped with: \emph{i)} an associative composition law defined as the concatenation of paths at the base point; \emph{ii)} an identity, defined as the class of (trivial) paths homotopic the base point; and \emph{iii)} an inverse, defined for each class by its paths travelled backwards. The fundamental group of $\mathbb{T}^{3N}$ is a free Abelian group of rank $3N$ and it is thus isomorphic to $\mathbb{Z}^{3N}$ \cite{FundGroup}. Therefore, given a base point $\barR(0) \in \mathbb{T}^{3N}$,
topologically equivalent loops can be \textit{uniquely} identified by the $3N$- integer tuple $\mathbf{n} = \{ n_{i\alpha} \}$, where $n_{i\alpha}$ is the \textit{winding number} of the $i$-th atom along the $\alpha$-th spatial direction. This is illustrated in Figure \ref{fig:ciambella} (right) in the toy case where the ACS has dimension 2 and the loop $\barC_I$ is represented by $\mathbf{n}=(1,1)$. Notice that with {a} such representation the concatenation of two loops $\barC_\mathbf{n} \circ \barC_\mathbf{m}$
is simply expressed as {the} sum of two integer vectors: $\mathbf{n}+\mathbf{m}= \{n_{i\alpha}+m_{i\alpha}\}$. Likewise, trivial loops are characterised by $\mathbf{n}=0$. In the following we assume that all trivial loops can be shrunk to a point without {ever closing the electronic gap}; this condition will be referred to in the following as \emph{strong adiabaticity}. As a generic loop $\barC_I$ is the concatenation of elementary loops involving individual atoms along specific directions, $\barC_{i\alpha}$, and the dipole displaced along each of them is likewise additive, we conclude that:
\begin{equation}
    Q_\alpha[\barC_I] = \sum_{i\beta}  q_{i\alpha \beta} \, n_{i\beta}, \label{eq:totQ_Cartesian}
\end{equation}
where $q_{i\alpha \beta} = Q_\alpha[\barC_{i\beta}]$ is the integer charge associated with the $\alpha$-th component of the dipole displaced by a loop of the $i$-th atom along the $\beta$-th direction, according to Eq.~\eqref{eq:Thouless}. Whenever the positions of two identical atoms can be interchanged without closing the electronic gap and strong adiabaticity holds, the dipole displaced along two trajectories that differ by such an {atomic} interchange coincide, and the topological charges $q_{i\alpha\beta}$ can only depend on $i$ through the species of the $i$-th atom, $S(i)$: $q_{i\alpha\beta} = q_{S(i),\alpha\beta}$. Also, the requirement that the dipole displaced along the sum of any two lattice vectors equals the sum of the dipoles displaced along each of them implies that the $q_S$ charges are (integer) scalars: $q_{S\alpha\beta}=q_S\delta_{\alpha\beta}$ \cite{Jiang2012}. We conclude that the dipole displaced along the $\barC_{II'}$ loop can be cast into the form:
\begin{equation}
        \Delta {\bm{\upmu}}_{II'} = \ell \sum_i q_{S(i)} \mathbf{n}_i, \label{eq:topoqdef}
\end{equation}
where $\mathbf{n}_i=\bigl ( n_{ix},n_{iy},n_{iz} \bigr )$ is the set of three winding numbers of the $i$-th atom in the $\barC_{II'}$ loop. The topological charges defined by Eq.~\eqref{eq:topoqdef} have all the properties that chemical common sense requires from oxidation numbers, and provide therefore a rigorous topological definition of them. Among the necessary but non-trivial consequences of this definition, we point out the additivity of the charge transported by several atoms that are being displaced simultaneously. This definition puts on a firm ground similar conclusions that could be drawn using the concept of Wannier centres \cite{Jiang2012}.

We now consider the dipole displacement computed from the $q_S$ topological charges:

\begin{align}
    \Delta{\bm{\upmu}}'(t)& = {\Omega} \int_0^t \mathbf{J}'(t')dt', \label{eq:DeltaP'}\\
     \mathbf{J}'(t) &= {\frac{1}{\Omega}}\sum_i q_{S(i)} \mathbf{v}_i(t). \label{eq:J'def}
\end{align}
Evidently, one has: $\Delta {\bm{\upmu}}'(t) = \Delta{\bm{\upmu}}_{II'} + \sum_i q_{S(i)}\int_{I'}^F d\mathbf{r}_i$. The second term on the right-hand side of this expression is bounded, and we conclude that:
\begin{equation}
    \lim_{t\to\infty}\frac{1}{t}\langle |\Delta{\bm{\upmu}}(t)|^2\rangle = \lim_{t\to\infty}\frac{1}{t}\langle |\Delta{\bm{\upmu}}'(t)|^2\rangle,
    \label{eq:EH-conclusiva}
\end{equation}
and therefore:
\begin{equation}
    \int_0^\infty \langle \mathbf J (t) \cdot \mathbf J (0) \rangle \, dt = \int_0^\infty \langle \mathbf J' (t) \cdot \mathbf J' (0) \rangle \, dt. \label{eq:equazione-conclusiva}
\end{equation}
Eqs.~(\ref{eq:EH-conclusiva}-\ref{eq:equazione-conclusiva}) are the main conclusion of our work: \emph{The adiabatic electrical conductivity of a liquid can be exactly obtained by replacing in Eq.~\eqref{eq:current} the time-dependent, real-valued, Born charge tensor of each atom with an integer, time-independent, scalar topological charge, which only depends on the atomic species, $q_{S(i)}$.}
The topological arguments in which this conclusion is rooted, while global and based on PBC by their very nature, naturally lead to the definition of such quantities as atomic oxidation numbers, which are both local and independent of the system size. This makes us believe that our conclusions hold in the thermodynamic limit and are independent of the boundary conditions being adopted.

The extent to which the above theory applies to molecular fluids, such as, \emph{e.g.}, ionic liquids, depends on the occurrence of one of the following two circumstances: \emph{i)} when a molecular species is stable in solution, \emph{i.e.} it does not coexist with any of its constituent moieties, our considerations show that the charge transported by it across a closed loop is quantised, and our conclusions hold under the same assumptions that are necessary in the atomic case; when a molecular species coexists with two or more of its constituent moieties, as it is the case, \emph{e.g.}, in partially ionic water \cite{French2011}, our considerations still hold under the hypothesis, which we may call \emph{adiabatic dissociation}, that the dissociation of a molecule into its constituent moieties occurs without closing the electronic gap. 

\begin{figure}
    \begin{center}
        \includegraphics[width=0.85\columnwidth]{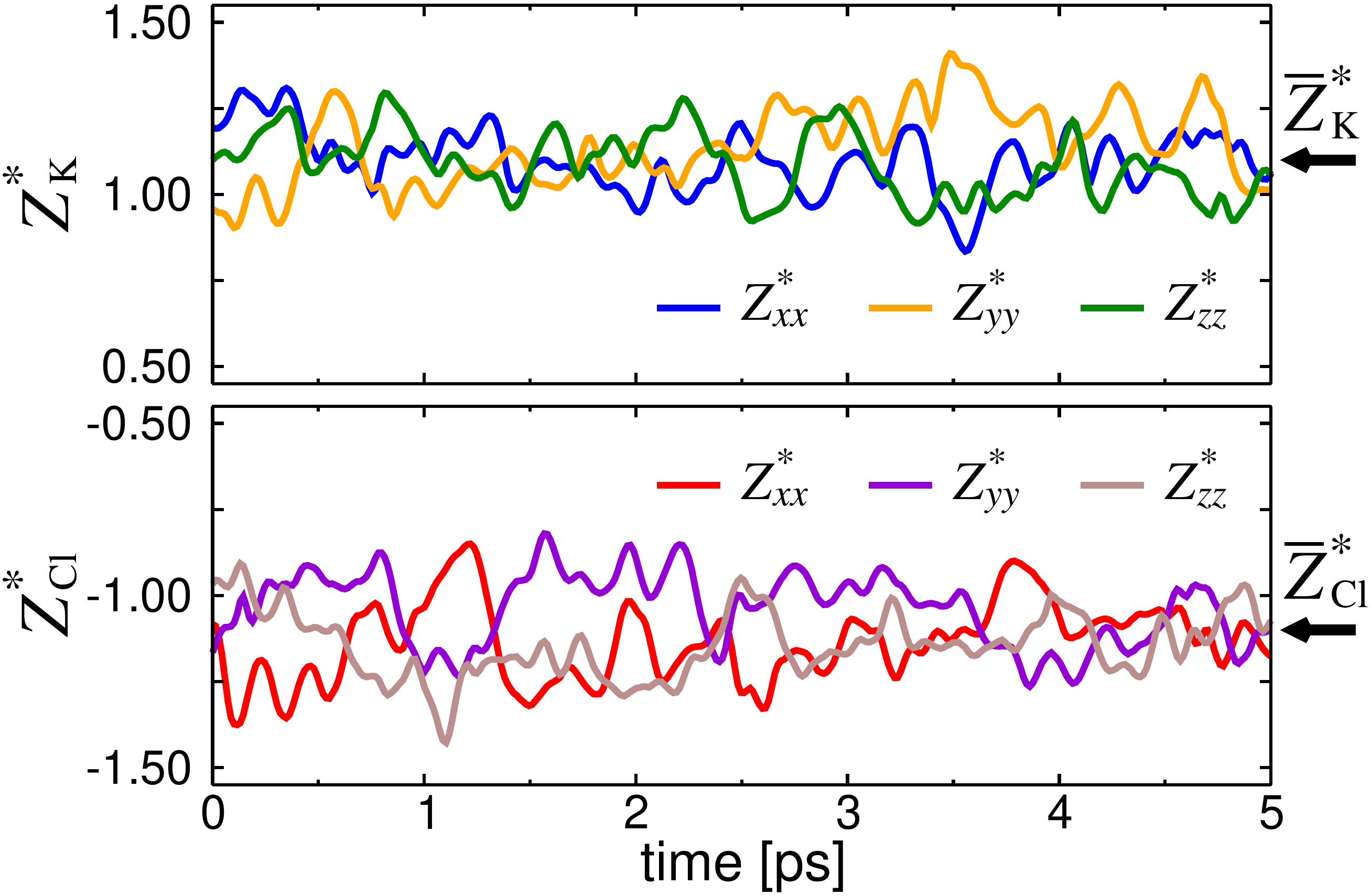}
        \caption{Time series of the Born effective-charge tensor. Shown are its three diagonal components for one K atom (top) and one Cl atom (bottom). The average values on the diagonal are: $\overline{Z}^*_\mathrm{K}=1.10 \pm 0.01$ and $\overline{Z}^*_\mathrm{Cl}=-1.10 \pm 0.01$, reported on the right.}
        \label{fig:ZeffKandCl}
    \end{center}
\end{figure}

\section{Numerical Experiments}

In order to demonstrate our results we have performed extensive numerical experiments on a 64-atom sample of molten KCl at a density of $1.42$ g/cm$^3$ \cite{Kirshenbaum1962}, corresponding to a cubic simulation cell whose edge is $\ell=14.07\,\mathrm{\AA}$. All simulations were performed using computer codes from the \textsc{Quantum Espresso} package v.6.1 \cite{Giannozzi2009,*Giannozzi2017}. 
We employed the PBE energy functional \cite{Perdew1996} with ONCV pseudopotentials \cite{Schlipf2015} and a plane-wave kinetic-energy cutoff of 55 Ry. AIMD simulations were performed within the Car-Parrinello method with a time step of 15 time a.u.~and a fictitious electronic mass of 400 electron masses.
We first ran a 90-ps AIMD simulation in the \emph{NVE} ensemble, following an \emph{NVT} thermalisation at 1200 K of a few ps, performed using a Nos\'e-Hoover-thermostat \cite{Nose1984,*Hoover1985}. The charge flux in Eq.~\eqref{eq:current} was sampled every 600 time a.u. ($\approx$ 14.5 fs) using Born effective-charge tensors, $\Zstar_i$, computed from density-functional perturbation theory \cite{Baroni2001}. In Figure \ref{fig:ZeffKandCl} we report a sample from the time series of the (diagonal elements of the) Born effective-charge tensors for a pair of K (top) and Cl (bottom) atoms. Notice the amplitude of the fluctuations around the average values that are reported on the right. The average effective charges are scalars because of overall rotational invariance and sum up to zero because of the acoustic sum rule. Note that, at variance with the topological charges / oxidation numbers defined above, the average effective charges are not integers ($|Z^*| = 1.10 \pm 0.01$).

\begin{figure}
    \centering
    \includegraphics[width = 0.8\linewidth]{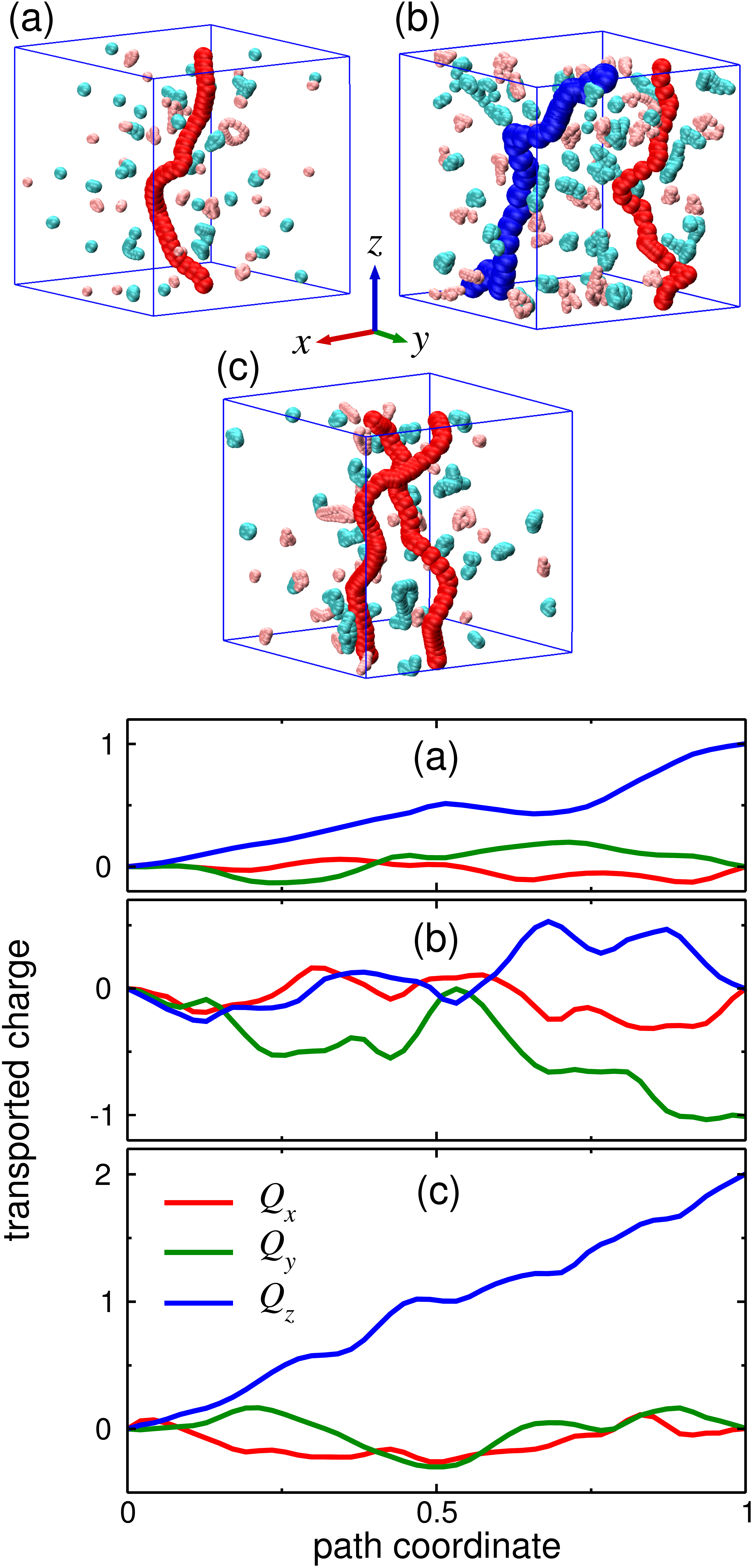}
    \caption{Closed paths and charge-transport quantisation. Upper panels: sample minimum-energy paths in a model of liquid KCl corresponding to the transport of a K ion by one unit cell along the $(001)$ direction (a); of a cation as before and an anion along the (011) direction (b); and two cations along (001) with an interchange between the two (c), see text. Lower panels: charge transported along the paths depicted above.}
    \label{fig:vermi_tri}
\end{figure}

We then calculated the topological charges, $q_S$, by evaluating the integral of Eq.~\eqref{eq:Thouless} along different paths whose end points are the same snapshot selected from our AIMD trajectory and such that only one or two atoms are transported by one unit cell. These paths were determined to be \emph{minimum energy paths} (MEP) and generated using the nudged elastic band method \cite{Jonsson1998}. All the atoms were let free to rearrange their positions along the MEP, as it is revealed by a close inspection of the figures, which show small but visible fluctuations in the positions of the atoms not participating in mass transport. Born effective-charge tensors were computed for every atom at each discrete image of the MEP{. The topological charges were} finally evaluated from the definition of displaced dipole: {$\Delta \mu_\alpha = \ell \sum_i q_{S(i)} n_{i\alpha} = \int d \mu_\alpha = \sum_{i\beta} \int Z^*_{i\alpha \beta} dr_{i\beta} $}, where $n_{i\alpha}$ is the winding number of the $i$-th atom in the $\alpha$-th Cartesian direction (\emph{cfr.~}Eq.~\ref{eq:topoqdef}). This is illustrated in Figure \ref{fig:vermi_tri}, where we display three such MEPs (top) and the corresponding charge transported along each of them (bottom). The (a) panels refer to the transport of a single K atom along the $z$ direction. Below we see that the charge transported along $z$ is $q_\mathrm{K}=1$, whereas that transported along $x$ or $y$ vanishes. In panels (b) one cation and one anion are transported. The anion moves from the cell conventionally labelled as at the origin, $\boldsymbol{\tau}^\circ=\ell(0,0,0)$, and the one located at $\boldsymbol{\tau}'=\ell(0,1,1)$ $\bigl ($winding numbers $\mathbf{n}_\mathrm{Cl}=(0,1,1)\bigr)$, whereas the cation moves from $\boldsymbol{\tau}^\circ$ to $\boldsymbol{\tau}''=\ell(0,0,1)$: the total charges transported along $z$ by the cation and the anion cancel exactly, whereas a net negative charge $q_\mathrm{Cl}=-1$ is transported by the anion along $y$. Finally, in panels (c) two cations are transported along $z$ \emph{and} interchanged. Topologically, this is not a loop, so that it seems that no conclusions about the total transported charge can be drawn. However, the concatenation of two such paths is indeed a loop with winding numbers $(0,0,2)$ for each transported cation, corresponding to a total transported charge $Q_z=4$. As the two open paths being concatenated are evidently equivalent, the charge transported along each of them is $Q_z=2$, as demonstrated in the (c) panel below. Panels (b) and (c) clearly show the additivity of topological charges.

\begin{figure}
    \begin{center}
        \includegraphics[width=0.9\columnwidth]{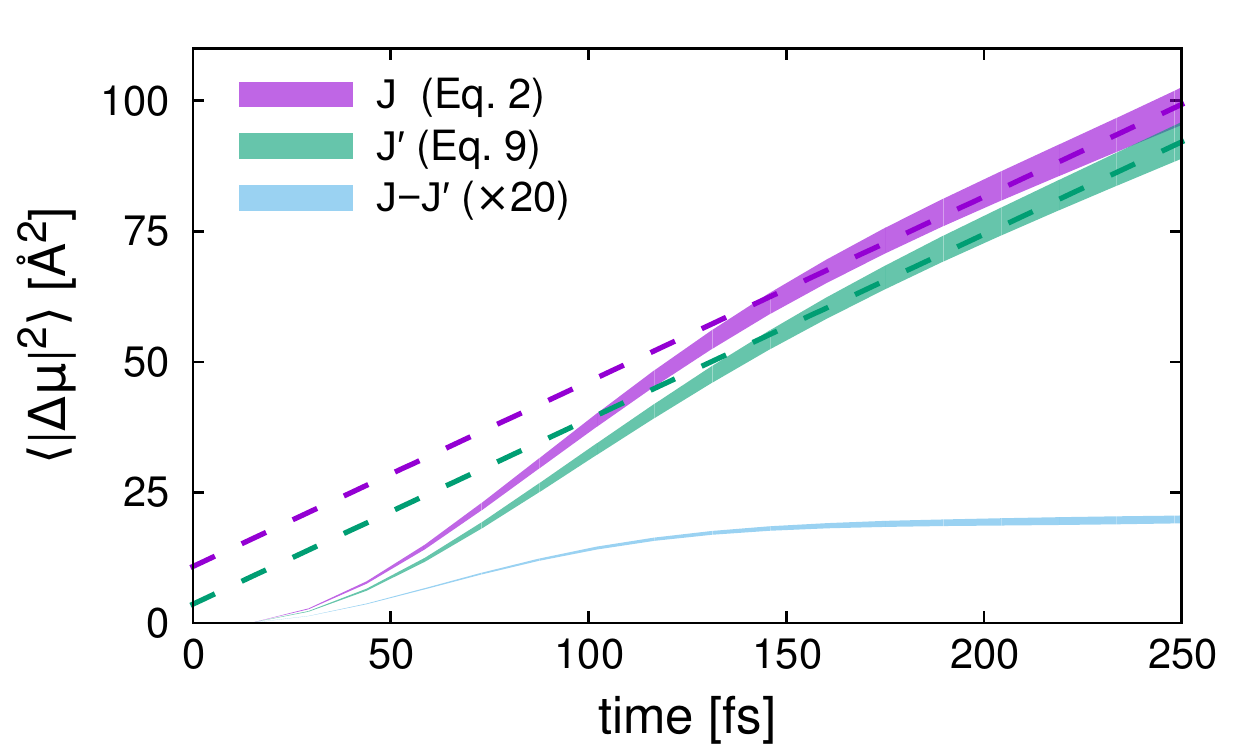}
        \caption{Dipole mean-square displacements vs time. The reported plots are computed using the charge flux defined in terms of the Born effective charges ($\mathbf{J}$, Eqs.~\ref{eq:current} and \ref{eq:DeltaP}, violet), in terms of topological charges ($\mathbf{J'}$, Eqs.~\ref{eq:J'def} and \ref{eq:DeltaP'}, light green), and using their difference (light blue). The width of the continuous lines indicates the estimated statistical error, while the dashed lines indicate a linear fit of the data at large time, indicating the diffusive behaviour.}
        \label{fig:ACF_EH}
    \end{center}
\end{figure}

In order to validate our final conclusions, Eqs.~(\ref{eq:EH-conclusiva},\ref{eq:equazione-conclusiva}), in Figure~\ref{fig:ACF_EH} we display the mean square dipole displaced by the {charge fluxes} $\mathbf{J}$ and $\mathbf{J}'$, Eqs. (\ref{eq:current},\ref{eq:DeltaP}) and (\ref{eq:DeltaP'},\ref{eq:J'def}), along with that displaced by their difference, $\mathbf{J}-\mathbf{J}'$, as functions of time. The two functions slightly, but markedly, differ, and their large-time slopes, which are proportional to the electrical conductivity, visibly coincide. Even more conspicuously, the dipole displaced by their difference, $\int_0^t\bigl ( \mathbf{J}(t')-\mathbf{J}'(t') \bigr ) dt'$ is bounded at large times, thus not contributing to the conductivity. A more refined analysis based on the cepstral analysis of the time series of the charge fluxes \cite{Ercole2017,Baroni2018} gives the same value within the same statistical uncertainty ($\sigma= 3.2 \pm 0.2 ~[\mathrm{S/cm}]$) when computed from Born effective charges or topological charges, whereas using the average Born effective charges would result in an overestimate by $\approx20\%$. Although our simulation settings were thought for validation purposes and the system size is certainly too small for ultimate accuracy, this value compares fairly with the experimental datum ($\sigma= 2.6~[\mathrm{S/cm]}$) at an estimated average simulation temperature of $1230 \pm 90$ K \cite{Janz1968}. When comparing theory with experiment, an additional incertitude of $\pm 0.15~[\mathrm{S/cm}]$ on the theoretical datum should be considered, on account of the large temperature fluctuations due to the small system size. Additional systematic finite-size errors come from the use of PBC and are difficult to estimate without a size-scaling analysis, but they are irrelevant to the conclusions of the present paper.

\section{Conclusions}

We conclude by expressing our confidence that the results presented in this paper will have a strong impact on both computer simulations and fundamental research. On the more practical side, our findings will allow a considerable simplification of the quantum numerical modelling of ionic conduction in complex systems and materials such as, \emph{e.g.}, ionic liquids \cite{Armand:2009is} or solid-state electrolytes \cite{Marcolongo2017,*Kahle2018}, avoiding, in many cases, the cumbersome and time-consuming computation of Born effective charges. On a more fundamental side, our work provides a solid topological foundation and generalisation to liquids of the definition of ionic oxidation states already available for solids \cite{Jiang2012}. This foundation will hopefully allow one to explore the limits of this definition and generalisation and their impact on ionic transport. For instance, in many systems the same ion is present in different oxidation states. As an important example, iron appears in both its ferrous and ferric ionic forms in water solution and in many oxides. Our analysis shows that the coexistence of different oxidation states for the same element in the same system may be due to the existence of {zero-gap} domains in the atomic configuration space that would be crossed by any atomic paths interchanging the positions of two identical ions in different oxidation states. While this scenario is likely the most common to occur, a different, more exotic, one cannot be excluded on purely topological grounds and its existence is worth exploring. In fact, when strong adiabaticity breaks, it is possible that two loops with the same winding numbers could not be distorted into one another without closing the electronic gap, and they may thus transport different, yet \textit{integer}, charges. While in the first scenario {closing the electronic gap} while swapping two like atoms would simply determine the chemically \emph{acceptable} inequivalence of the oxidation numbers of two identical atoms in different local environments, the second scenario would imply the chemically wicked situation where two different oxidation states can be attached to \emph{the same atom in the same local environment}. As a consequence, one could observe a non-vanishing adiabatic charge transport without a net mass transport \cite{[{See the discussion at pp.~51--52 in }]Resta-Vanderbilt2007}.

\section{Acknowledgements}

\begin{acknowledgments} We are grateful to Raffaele Resta for insightful discussions and to Riccardo Bertossa for technical assistance. This work was partially funded by the EU through the \textsc{MaX} Centre of Excellence for supercomputing applications (Projects No.~676598 and 824143). \end{acknowledgments}

\section{Author Contributions}
Both authors contributed to all aspects of this work.

\section{Competing Interests statement}
The authors declare no competing financial interests.

\section{Data Availability statement}
The data that support the plots within this paper and other findings of this study are available from the corresponding author upon reasonable request.


\begin{thebibliography}{33}%
\makeatletter
\providecommand \@ifxundefined [1]{%
 \@ifx{#1\undefined}
}%
\providecommand \@ifnum [1]{%
 \ifnum #1\expandafter \@firstoftwo
 \else \expandafter \@secondoftwo
 \fi
}%
\providecommand \@ifx [1]{%
 \ifx #1\expandafter \@firstoftwo
 \else \expandafter \@secondoftwo
 \fi
}%
\providecommand \natexlab [1]{#1}%
\providecommand \enquote  [1]{``#1''}%
\providecommand \bibnamefont  [1]{#1}%
\providecommand \bibfnamefont [1]{#1}%
\providecommand \citenamefont [1]{#1}%
\providecommand \href@noop [0]{\@secondoftwo}%
\providecommand \href [0]{\begingroup \@sanitize@url \@href}%
\providecommand \@href[1]{\@@startlink{#1}\@@href}%
\providecommand \@@href[1]{\endgroup#1\@@endlink}%
\providecommand \@sanitize@url [0]{\catcode `\\12\catcode `\$12\catcode
  `\&12\catcode `\#12\catcode `\^12\catcode `\_12\catcode `\%12\relax}%
\providecommand \@@startlink[1]{}%
\providecommand \@@endlink[0]{}%
\providecommand \url  [0]{\begingroup\@sanitize@url \@url }%
\providecommand \@url [1]{\endgroup\@href {#1}{\urlprefix }}%
\providecommand \urlprefix  [0]{URL }%
\providecommand \Eprint [0]{\href }%
\providecommand \doibase [0]{http://dx.doi.org/}%
\providecommand \selectlanguage [0]{\@gobble}%
\providecommand \bibinfo  [0]{\@secondoftwo}%
\providecommand \bibfield  [0]{\@secondoftwo}%
\providecommand \translation [1]{[#1]}%
\providecommand \BibitemOpen [0]{}%
\providecommand \bibitemStop [0]{}%
\providecommand \bibitemNoStop [0]{.\EOS\space}%
\providecommand \EOS [0]{\spacefactor3000\relax}%
\providecommand \BibitemShut  [1]{\csname bibitem#1\endcsname}%
\let\auto@bib@innerbib\@empty
\bibitem [{\citenamefont {Green}(1952)}]{Green1952}%
  \BibitemOpen
  \bibfield  {author} {\bibinfo {author} {\bibfnamefont {MS}~\bibnamefont
  {Green}},\ }\bibfield  {title} {\enquote {\bibinfo {title} {Markoff random
  processes and the statistical mechanics of time‐dependent phenomena.}}\
  }\href {\doibase 10.1063/1.1700722} {\bibfield  {journal} {\bibinfo
  {journal} {J. Chem. Phys.}\ }\textbf {\bibinfo {volume} {20}},\ \bibinfo
  {pages} {1281--1295} (\bibinfo {year} {1952})}\BibitemShut {NoStop}%
\bibitem [{\citenamefont {Green}(1954)}]{Green1954}%
  \BibitemOpen
  \bibfield  {author} {\bibinfo {author} {\bibfnamefont {MS}~\bibnamefont
  {Green}},\ }\bibfield  {title} {\enquote {\bibinfo {title} {Markoff random
  processes and the statistical mechanics of time-dependent phenomena. ii.
  irreversible processes in fluids},}\ }\href {\doibase 10.1063/1.1740082}
  {\bibfield  {journal} {\bibinfo  {journal} {J. Chem. Phys.}\ }\textbf
  {\bibinfo {volume} {22}},\ \bibinfo {pages} {398--413} (\bibinfo {year}
  {1954})}\BibitemShut {NoStop}%
\bibitem [{\citenamefont {Kubo}(1957)}]{Kubo1957a}%
  \BibitemOpen
  \bibfield  {author} {\bibinfo {author} {\bibfnamefont {R}~\bibnamefont
  {Kubo}},\ }\bibfield  {title} {\enquote {\bibinfo {title}
  {Statistical-mechanical theory of irreversible processes. i. {General} theory
  and simple applications to magnetic and conduction problems},}\ }\href
  {\doibase 10.1143/JPSJ.12.570} {\bibfield  {journal} {\bibinfo  {journal} {J.
  Phys. Soc. Jpn.}\ }\textbf {\bibinfo {volume} {12}},\ \bibinfo {pages}
  {570--586} (\bibinfo {year} {1957})}\BibitemShut {NoStop}%
\bibitem [{\citenamefont {Kubo}\ \emph {et~al.}(1957)\citenamefont {Kubo},
  \citenamefont {Yokota},\ and\ \citenamefont {Nakajima}}]{Kubo1957b}%
  \BibitemOpen
  \bibfield  {author} {\bibinfo {author} {\bibfnamefont {R}~\bibnamefont
  {Kubo}}, \bibinfo {author} {\bibfnamefont {M}~\bibnamefont {Yokota}}, \ and\
  \bibinfo {author} {\bibfnamefont {S}~\bibnamefont {Nakajima}},\ }\bibfield
  {title} {\enquote {\bibinfo {title} {Statistical-mechanical theory of
  irreversible processes. ii. response to thermal disturbance},}\ }\href
  {\doibase 10.1143/JPSJ.12.1203} {\bibfield  {journal} {\bibinfo  {journal}
  {J. Phys. Soc. Jpn.}\ }\textbf {\bibinfo {volume} {12}},\ \bibinfo {pages}
  {1203--1211} (\bibinfo {year} {1957})}\BibitemShut {NoStop}%
\bibitem [{\citenamefont {Baroni}\ \emph {et~al.}(2001)\citenamefont {Baroni},
  \citenamefont {de~Gironcoli}, \citenamefont {{Dal Corso}},\ and\
  \citenamefont {Giannozzi}}]{Baroni2001}%
  \BibitemOpen
  \bibfield  {author} {\bibinfo {author} {\bibfnamefont {S}~\bibnamefont
  {Baroni}}, \bibinfo {author} {\bibfnamefont {S}~\bibnamefont {de~Gironcoli}},
  \bibinfo {author} {\bibfnamefont {A}~\bibnamefont {{Dal Corso}}}, \ and\
  \bibinfo {author} {\bibfnamefont {P}~\bibnamefont {Giannozzi}},\ }\bibfield
  {title} {\enquote {\bibinfo {title} {Phonons and related crystal properties
  from density-functional perturbation theory},}\ }\href {\doibase
  10.1103/RevModPhys.73.515} {\bibfield  {journal} {\bibinfo  {journal} {Rev.
  Mod. Phys.}\ }\textbf {\bibinfo {volume} {73}},\ \bibinfo {pages} {515--562}
  (\bibinfo {year} {2001})}\BibitemShut {NoStop}%
\bibitem [{\citenamefont {Resta}(2010)}]{Resta2010}%
  \BibitemOpen
  \bibfield  {author} {\bibinfo {author} {\bibfnamefont {R}~\bibnamefont
  {Resta}},\ }\bibfield  {title} {\enquote {\bibinfo {title} {{Electrical
  polarization and orbital magnetization: The modern theories}},}\ }\href
  {\doibase 10.1088/0953-8984/22/12/123201} {\bibfield  {journal} {\bibinfo
  {journal} {J. Phys. Condens. Matter}\ }\textbf {\bibinfo {volume} {22}},\
  \bibinfo {pages} {123201} (\bibinfo {year} {2010})}\BibitemShut {NoStop}%
\bibitem [{\citenamefont {Vanderbilt}(2018)}]{Vanderbilt2018}%
  \BibitemOpen
  \bibfield  {author} {\bibinfo {author} {\bibfnamefont {D}~\bibnamefont
  {Vanderbilt}},\ }\href@noop {} {\emph {\bibinfo {title} {Berry Phases in
  Electronic Structure Theory: Electric Polarization, Orbital Magnetization and
  Topological Insulators}}}\ (\bibinfo  {publisher} {Cambridge University
  Press},\ \bibinfo {year} {2018})\BibitemShut {NoStop}%
\bibitem [{\citenamefont {French}\ \emph {et~al.}(2011)\citenamefont {French},
  \citenamefont {Hamel},\ and\ \citenamefont {Redmer}}]{French2011}%
  \BibitemOpen
  \bibfield  {author} {\bibinfo {author} {\bibfnamefont {Martin}\ \bibnamefont
  {French}}, \bibinfo {author} {\bibfnamefont {Sebastien}\ \bibnamefont
  {Hamel}}, \ and\ \bibinfo {author} {\bibfnamefont {Ronald}\ \bibnamefont
  {Redmer}},\ }\bibfield  {title} {\enquote {\bibinfo {title} {{Dynamical
  Screening and Ionic Conductivity in Water from Ab Initio Simulations}},}\
  }\href {\doibase 10.1103/PhysRevLett.107.185901} {\bibfield  {journal}
  {\bibinfo  {journal} {Phys. Rev. Lett.}\ }\textbf {\bibinfo {volume} {107}},\
  \bibinfo {pages} {185901} (\bibinfo {year} {2011})}\BibitemShut {NoStop}%
\bibitem [{\citenamefont {Jiang}\ \emph {et~al.}(2012)\citenamefont {Jiang},
  \citenamefont {Levchenko},\ and\ \citenamefont {Rappe}}]{Jiang2012}%
  \BibitemOpen
  \bibfield  {author} {\bibinfo {author} {\bibfnamefont {L}~\bibnamefont
  {Jiang}}, \bibinfo {author} {\bibfnamefont {SV}~\bibnamefont {Levchenko}}, \
  and\ \bibinfo {author} {\bibfnamefont {AM}~\bibnamefont {Rappe}},\ }\bibfield
   {title} {\enquote {\bibinfo {title} {{Rigorous definition of oxidation
  states of ions in solids}},}\ }\href {\doibase
  10.1103/PhysRevLett.108.166403} {\bibfield  {journal} {\bibinfo  {journal}
  {Phys. Rev. Lett.}\ }\textbf {\bibinfo {volume} {108}},\ \bibinfo {pages}
  {1--5} (\bibinfo {year} {2012})},\ \Eprint {http://arxiv.org/abs/1106.2836}
  {arXiv:1106.2836} \BibitemShut {NoStop}%
\bibitem [{\citenamefont {Marcolongo}\ \emph {et~al.}(2016)\citenamefont
  {Marcolongo}, \citenamefont {Umari},\ and\ \citenamefont
  {Baroni}}]{Marcolongo2016}%
  \BibitemOpen
  \bibfield  {author} {\bibinfo {author} {\bibfnamefont {A}~\bibnamefont
  {Marcolongo}}, \bibinfo {author} {\bibfnamefont {P}~\bibnamefont {Umari}}, \
  and\ \bibinfo {author} {\bibfnamefont {S}~\bibnamefont {Baroni}},\ }\bibfield
   {title} {\enquote {\bibinfo {title} {Microscopic theory and ab initio
  simulation of atomic heat transport},}\ }\href {\doibase 10.1038/nphys3509}
  {\bibfield  {journal} {\bibinfo  {journal} {Nature Phys.}\ }\textbf {\bibinfo
  {volume} {12}},\ \bibinfo {pages} {80--84} (\bibinfo {year}
  {2016})}\BibitemShut {NoStop}%
\bibitem [{\citenamefont {Ercole}\ \emph {et~al.}(2016)\citenamefont {Ercole},
  \citenamefont {Marcolongo}, \citenamefont {Umari},\ and\ \citenamefont
  {Baroni}}]{Ercole2016}%
  \BibitemOpen
  \bibfield  {author} {\bibinfo {author} {\bibfnamefont {L}~\bibnamefont
  {Ercole}}, \bibinfo {author} {\bibfnamefont {A}~\bibnamefont {Marcolongo}},
  \bibinfo {author} {\bibfnamefont {P}~\bibnamefont {Umari}}, \ and\ \bibinfo
  {author} {\bibfnamefont {S}~\bibnamefont {Baroni}},\ }\bibfield  {title}
  {\enquote {\bibinfo {title} {Gauge invariance of thermal transport
  coefficients},}\ }\href {\doibase 10.1007/s10909-016-1617-6} {\bibfield
  {journal} {\bibinfo  {journal} {J. Low Temp. Phys.}\ }\textbf {\bibinfo
  {volume} {185}},\ \bibinfo {pages} {79--86} (\bibinfo {year}
  {2016})}\BibitemShut {NoStop}%
\bibitem [{\citenamefont {Baroni}\ \emph {et~al.}(2018)\citenamefont {Baroni},
  \citenamefont {Bertossa}, \citenamefont {Ercole}, \citenamefont {Grasselli},\
  and\ \citenamefont {Marcolongo}}]{Baroni2018}%
  \BibitemOpen
  \bibfield  {author} {\bibinfo {author} {\bibfnamefont {Stefano}\ \bibnamefont
  {Baroni}}, \bibinfo {author} {\bibfnamefont {Riccardo}\ \bibnamefont
  {Bertossa}}, \bibinfo {author} {\bibfnamefont {Loris}\ \bibnamefont
  {Ercole}}, \bibinfo {author} {\bibfnamefont {Federico}\ \bibnamefont
  {Grasselli}}, \ and\ \bibinfo {author} {\bibfnamefont {Aris}\ \bibnamefont
  {Marcolongo}},\ }\enquote {\bibinfo {title} {Heat transport in insulators
  from ab initio {G}reen-{K}ubo theory},}\ in\ \href {\doibase
  10.1007/978-3-319-50257-1_12-1} {\emph {\bibinfo {booktitle} {Handbook of
  Materials Modeling: Applications: Current and Emerging Materials}}},\
  \bibinfo {editor} {edited by\ \bibinfo {editor} {\bibfnamefont {Wanda}\
  \bibnamefont {Andreoni}}\ and\ \bibinfo {editor} {\bibfnamefont {Sidney}\
  \bibnamefont {Yip}}}\ (\bibinfo  {publisher} {Springer International
  Publishing},\ \bibinfo {address} {Cham},\ \bibinfo {year} {2018})\ pp.\
  \bibinfo {pages} {1--36},\ \bibinfo {edition} {2nd}\ ed.,\ \Eprint
  {http://arxiv.org/abs/1802.08006} {arXiv:1802.08006 [cond-mat.stat-mech]}
  \BibitemShut {NoStop}%
\bibitem [{\citenamefont {Helfand}(1960)}]{Helfand1960}%
  \BibitemOpen
  \bibfield  {author} {\bibinfo {author} {\bibfnamefont {E}~\bibnamefont
  {Helfand}},\ }\bibfield  {title} {\enquote {\bibinfo {title} {Transport
  coefficients from dissipation in a canonical ensemble},}\ }\href {\doibase
  10.1103/PhysRev.119.1} {\bibfield  {journal} {\bibinfo  {journal} {Phys.
  Rev.}\ }\textbf {\bibinfo {volume} {119}},\ \bibinfo {pages} {1--9} (\bibinfo
  {year} {1960})}\BibitemShut {NoStop}%
\bibitem [{\citenamefont {Marcolongo}(2014)}]{Marcolongo2014}%
  \BibitemOpen
  \bibfield  {author} {\bibinfo {author} {\bibfnamefont {A}~\bibnamefont
  {Marcolongo}},\ }\emph {\bibinfo {title} {Theory and ab initio simulation of
  atomic heat transport}},\ \href {https://cm.sissa.it/thesis/2014/marcolongo}
  {Ph.D. thesis},\ \bibinfo  {school} {Scuola Internazionale Superiore di Studi
  Avanzati}, \bibinfo {address} {Trieste} (\bibinfo {year} {2014}),\ \bibinfo
  {note} {\url{https://cm.sissa.it/thesis/2014/marcolongo}}\BibitemShut
  {NoStop}%
\bibitem [{\citenamefont {Resta}(2017)}]{RestaJuelich}%
  \BibitemOpen
  \bibfield  {author} {\bibinfo {author} {\bibfnamefont {Raffaele}\
  \bibnamefont {Resta}},\ }\enquote {\bibinfo {title} {The insulating state of
  matter: A geometrical theory},}\ in\ \href
  {https://www.cond-mat.de/events/correl17/manuscripts/resta.pdf} {\emph
  {\bibinfo {booktitle} {The Physics of Correlated Insulators, Metals, and
  Superconductors. Modeling and Simulation, Vol. 7}}},\ \bibinfo {editor}
  {edited by\ \bibinfo {editor} {\bibfnamefont {Eva}\ \bibnamefont {Pavarini}},
  \bibinfo {editor} {\bibfnamefont {Erik}\ \bibnamefont {Koch}}, \bibinfo
  {editor} {\bibfnamefont {Richard}\ \bibnamefont {Scalettar}}, \ and\ \bibinfo
  {editor} {\bibfnamefont {Richard~M.}\ \bibnamefont {Martin}}}\ (\bibinfo
  {publisher} {Verlag des Forschungszentrum Jülich},\ \bibinfo {year} {2017})\
  p.\ \bibinfo {pages} {3.5}\BibitemShut {NoStop}%
\bibitem [{\citenamefont {Thouless}(1983)}]{Thouless1983}%
  \BibitemOpen
  \bibfield  {author} {\bibinfo {author} {\bibfnamefont {DJ}~\bibnamefont
  {Thouless}},\ }\bibfield  {title} {\enquote {\bibinfo {title} {Quantization
  of particle transport},}\ }\href@noop {} {\bibfield  {journal} {\bibinfo
  {journal} {Phys. Rev. B}\ }\textbf {\bibinfo {volume} {27}},\ \bibinfo
  {pages} {6083--6087} (\bibinfo {year} {1983})}\BibitemShut {NoStop}%
\bibitem [{\citenamefont {Pendry}\ and\ \citenamefont
  {Hodges}(1984)}]{Pendry1984}%
  \BibitemOpen
  \bibfield  {author} {\bibinfo {author} {\bibfnamefont {J~B}\ \bibnamefont
  {Pendry}}\ and\ \bibinfo {author} {\bibfnamefont {C~H}\ \bibnamefont
  {Hodges}},\ }\bibfield  {title} {\enquote {\bibinfo {title} {The quantisation
  of charge transport in ionic systems},}\ }\href {\doibase
  10.1088/0022-3719/17/7/019} {\bibfield  {journal} {\bibinfo  {journal} {J.
  Phys. C}\ }\textbf {\bibinfo {volume} {17}},\ \bibinfo {pages} {1269--1279}
  (\bibinfo {year} {1984})}\BibitemShut {NoStop}%
\bibitem [{\citenamefont {Resta}(1998)}]{RESTA1998}%
  \BibitemOpen
  \bibfield  {author} {\bibinfo {author} {\bibfnamefont {R}~\bibnamefont
  {Resta}},\ }\bibfield  {title} {\enquote {\bibinfo {title}
  {{Quantum-mechanical position operator in extended systems}},}\ }\href@noop
  {} {\bibfield  {journal} {\bibinfo  {journal} {Physical Review Letters}\
  }\textbf {\bibinfo {volume} {80}},\ \bibinfo {pages} {1800--1803} (\bibinfo
  {year} {1998})}\BibitemShut {NoStop}%
\bibitem [{\citenamefont {Rowland}\ and\ \citenamefont
  {Weisstein}()}]{FundGroup}%
  \BibitemOpen
  \bibfield  {author} {\bibinfo {author} {\bibfnamefont {T}~\bibnamefont
  {Rowland}}\ and\ \bibinfo {author} {\bibfnamefont {EW}~\bibnamefont
  {Weisstein}},\ }\href {http://mathworld.wolfram.com/FundamentalGroup.html}
  {\enquote {\bibinfo {title} {Fundamental group},}\ }\bibinfo {note} {From
  MathWorld -- A Wolfram Web Resource,
  \url{http://mathworld.wolfram.com/FundamentalGroup.html}}\BibitemShut
  {NoStop}%
\bibitem [{\citenamefont {Kirshenbaum}\ \emph {et~al.}(1962)\citenamefont
  {Kirshenbaum}, \citenamefont {Cahill}, \citenamefont {McGonigal},\ and\
  \citenamefont {Grosse}}]{Kirshenbaum1962}%
  \BibitemOpen
  \bibfield  {author} {\bibinfo {author} {\bibfnamefont {AD}~\bibnamefont
  {Kirshenbaum}}, \bibinfo {author} {\bibfnamefont {JA}~\bibnamefont {Cahill}},
  \bibinfo {author} {\bibfnamefont {PJ}~\bibnamefont {McGonigal}}, \ and\
  \bibinfo {author} {\bibfnamefont {AV}~\bibnamefont {Grosse}},\ }\bibfield
  {title} {\enquote {\bibinfo {title} {{The density of liquid \uppercase{N}aCl
  and \uppercase{K}Cl and an estimate of their critical constants together with
  those of the other alkali halides}},}\ }\href@noop {} {\bibfield  {journal}
  {\bibinfo  {journal} {J. Inorg. Nucl. Chem.}\ }\textbf {\bibinfo {volume}
  {24}},\ \bibinfo {pages} {1287--1296} (\bibinfo {year} {1962})}\BibitemShut
  {NoStop}%
\bibitem [{\citenamefont {Giannozzi}\ \emph {et~al.}(2009)\citenamefont
  {Giannozzi} \emph {et~al.}}]{Giannozzi2009}%
  \BibitemOpen
  \bibfield  {author} {\bibinfo {author} {\bibfnamefont {P.}~\bibnamefont
  {Giannozzi}} \emph {et~al.},\ }\bibfield  {title} {\enquote {\bibinfo {title}
  {{QUANTUM ESPRESSO}: a modular and open-source software project for quantum
  simulations of materials},}\ }\href {\doibase 10.1088/0953-8984/21/39/395502}
  {\bibfield  {journal} {\bibinfo  {journal} {J. Phys. Condens. Matter}\
  }\textbf {\bibinfo {volume} {21}},\ \bibinfo {pages} {395502} (\bibinfo
  {year} {2009})}\BibitemShut {NoStop}%
\bibitem [{\citenamefont {Giannozzi}\ \emph {et~al.}(2017)\citenamefont
  {Giannozzi} \emph {et~al.}}]{Giannozzi2017}%
  \BibitemOpen
  \bibfield  {author} {\bibinfo {author} {\bibfnamefont {P.}~\bibnamefont
  {Giannozzi}} \emph {et~al.},\ }\bibfield  {title} {\enquote {\bibinfo {title}
  {Advanced capabilities for materials modelling with quantum espresso},}\
  }\href@noop {} {\bibfield  {journal} {\bibinfo  {journal} {J. Phys. Condens.
  Matter}\ }\textbf {\bibinfo {volume} {29}},\ \bibinfo {pages} {465901}
  (\bibinfo {year} {2017})}\BibitemShut {NoStop}%
\bibitem [{\citenamefont {Perdew}\ \emph {et~al.}(1996)\citenamefont {Perdew},
  \citenamefont {Burke},\ and\ \citenamefont {Ernzerhof}}]{Perdew1996}%
  \BibitemOpen
  \bibfield  {author} {\bibinfo {author} {\bibfnamefont {JP}~\bibnamefont
  {Perdew}}, \bibinfo {author} {\bibfnamefont {K}~\bibnamefont {Burke}}, \ and\
  \bibinfo {author} {\bibfnamefont {M}~\bibnamefont {Ernzerhof}},\ }\bibfield
  {title} {\enquote {\bibinfo {title} {Generalized gradient approximation made
  simple},}\ }\href {\doibase 10.1103/PhysRevLett.77.3865} {\bibfield
  {journal} {\bibinfo  {journal} {Phys. Rev. Lett.}\ }\textbf {\bibinfo
  {volume} {77}},\ \bibinfo {pages} {3865--3868} (\bibinfo {year}
  {1996})}\BibitemShut {NoStop}%
\bibitem [{\citenamefont {Schlipf}\ and\ \citenamefont
  {Gygi}(2015)}]{Schlipf2015}%
  \BibitemOpen
  \bibfield  {author} {\bibinfo {author} {\bibfnamefont {M}~\bibnamefont
  {Schlipf}}\ and\ \bibinfo {author} {\bibfnamefont {F}~\bibnamefont {Gygi}},\
  }\bibfield  {title} {\enquote {\bibinfo {title} {Optimization algorithm for
  the generation of oncv pseudopotentials},}\ }\href {\doibase
  https://doi.org/10.1016/j.cpc.2015.05.011} {\bibfield  {journal} {\bibinfo
  {journal} {Computer Physics Communications}\ }\textbf {\bibinfo {volume}
  {196}},\ \bibinfo {pages} {36 -- 44} (\bibinfo {year} {2015})},\ \bibinfo
  {note} {with pseudopotentials downloaded from
  \url{http://www.quantum-simulation.org/potentials/sg15_oncv/upf/}}\BibitemShut
  {NoStop}%
\bibitem [{\citenamefont {Nos{\'e}}(1984)}]{Nose1984}%
  \BibitemOpen
  \bibfield  {author} {\bibinfo {author} {\bibfnamefont {S}~\bibnamefont
  {Nos{\'e}}},\ }\bibfield  {title} {\enquote {\bibinfo {title} {A unified
  formulation of the constant temperature molecular dynamics methods},}\
  }\href@noop {} {\bibfield  {journal} {\bibinfo  {journal} {J. Chem. Phys.}\
  }\textbf {\bibinfo {volume} {81}},\ \bibinfo {pages} {511--519} (\bibinfo
  {year} {1984})}\BibitemShut {NoStop}%
\bibitem [{\citenamefont {Hoover}(1985)}]{Hoover1985}%
  \BibitemOpen
  \bibfield  {author} {\bibinfo {author} {\bibfnamefont {WG}~\bibnamefont
  {Hoover}},\ }\bibfield  {title} {\enquote {\bibinfo {title} {Canonical
  dynamics: equilibrium phase-space distributions},}\ }\href@noop {} {\bibfield
   {journal} {\bibinfo  {journal} {Phys. Rev. A}\ }\textbf {\bibinfo {volume}
  {31}},\ \bibinfo {pages} {1695--1697} (\bibinfo {year} {1985})}\BibitemShut
  {NoStop}%
\bibitem [{\citenamefont {J\'onsson}\ \emph {et~al.}(1998)\citenamefont
  {J\'onsson}, \citenamefont {Mills},\ and\ \citenamefont
  {Jacobsen}}]{Jonsson1998}%
  \BibitemOpen
  \bibfield  {author} {\bibinfo {author} {\bibfnamefont {H.}~\bibnamefont
  {J\'onsson}}, \bibinfo {author} {\bibfnamefont {G.}~\bibnamefont {Mills}}, \
  and\ \bibinfo {author} {\bibfnamefont {K.~W.}\ \bibnamefont {Jacobsen}},\
  }\enquote {\bibinfo {title} {Nudged elastic band method for finding minimum
  energy paths of transitions},}\ in\ \href@noop {} {\emph {\bibinfo
  {booktitle} {Classical and Quantum Dynamics in Condensed Phase
  Simulations}}},\ \bibinfo {editor} {edited by\ \bibinfo {editor}
  {\bibfnamefont {B.~J.}\ \bibnamefont {Berne}}, \bibinfo {editor}
  {\bibfnamefont {G.}~\bibnamefont {Ciccotti}}, \ and\ \bibinfo {editor}
  {\bibfnamefont {D.~F.}\ \bibnamefont {Coker}}}\ (\bibinfo  {publisher} {World
  Scientific},\ \bibinfo {year} {1998})\ p.\ \bibinfo {pages}
  {385–404}\BibitemShut {NoStop}%
\bibitem [{\citenamefont {Ercole}\ \emph {et~al.}(2017)\citenamefont {Ercole},
  \citenamefont {Marcolongo},\ and\ \citenamefont {Baroni}}]{Ercole2017}%
  \BibitemOpen
  \bibfield  {author} {\bibinfo {author} {\bibfnamefont {L}~\bibnamefont
  {Ercole}}, \bibinfo {author} {\bibfnamefont {A}~\bibnamefont {Marcolongo}}, \
  and\ \bibinfo {author} {\bibfnamefont {S}~\bibnamefont {Baroni}},\ }\bibfield
   {title} {\enquote {\bibinfo {title} {Accurate thermal conductivities from
  optimally short molecular dynamics simulations},}\ }\href {\doibase
  10.1038/s41598-017-15843-2} {\bibfield  {journal} {\bibinfo  {journal} {Sci.
  Rep.}\ }\textbf {\bibinfo {volume} {7}},\ \bibinfo {pages} {15835} (\bibinfo
  {year} {2017})}\BibitemShut {NoStop}%
\bibitem [{\citenamefont {Janz}\ \emph {et~al.}(1968)\citenamefont {Janz},
  \citenamefont {Dampier}, \citenamefont {Lakshminarayanan}, \citenamefont
  {Lorenz},\ and\ \citenamefont {Tomkins}}]{Janz1968}%
  \BibitemOpen
  \bibfield  {author} {\bibinfo {author} {\bibfnamefont {G.~J.}\ \bibnamefont
  {Janz}}, \bibinfo {author} {\bibfnamefont {F.~W.}\ \bibnamefont {Dampier}},
  \bibinfo {author} {\bibfnamefont {G.~R.}\ \bibnamefont {Lakshminarayanan}},
  \bibinfo {author} {\bibfnamefont {P.~K.}\ \bibnamefont {Lorenz}}, \ and\
  \bibinfo {author} {\bibfnamefont {R.~P.~T.}\ \bibnamefont {Tomkins}},\
  }\href@noop {} {\emph {\bibinfo {title} {Molten Salts: Volume I. Electrical
  Conductance, Density, and Viscosity Data.}}}\ (\bibinfo  {publisher} {U.S.
  National Bureau of Standards},\ \bibinfo {year} {1968})\ p.~\bibinfo {pages}
  {48}\BibitemShut {NoStop}%
\bibitem [{\citenamefont {Armand}\ \emph {et~al.}(2009)\citenamefont {Armand},
  \citenamefont {Endres}, \citenamefont {MacFarlane}, \citenamefont {Ohno},\
  and\ \citenamefont {Scrosati}}]{Armand:2009is}%
  \BibitemOpen
  \bibfield  {author} {\bibinfo {author} {\bibfnamefont {M}~\bibnamefont
  {Armand}}, \bibinfo {author} {\bibfnamefont {F}~\bibnamefont {Endres}},
  \bibinfo {author} {\bibfnamefont {Douglas~R}\ \bibnamefont {MacFarlane}},
  \bibinfo {author} {\bibfnamefont {H}~\bibnamefont {Ohno}}, \ and\ \bibinfo
  {author} {\bibfnamefont {B}~\bibnamefont {Scrosati}},\ }\bibfield  {title}
  {\enquote {\bibinfo {title} {{Ionic-liquid materials for the electrochemical
  challenges of the future}},}\ }\href {\doibase 10.1038/nmat2448} {\bibfield
  {journal} {\bibinfo  {journal} {Nat. Mater.}\ }\textbf {\bibinfo {volume}
  {8}},\ \bibinfo {pages} {621--629} (\bibinfo {year} {2009})}\BibitemShut
  {NoStop}%
\bibitem [{\citenamefont {Marcolongo}\ and\ \citenamefont
  {Marzari}(2017)}]{Marcolongo2017}%
  \BibitemOpen
  \bibfield  {author} {\bibinfo {author} {\bibfnamefont {A}~\bibnamefont
  {Marcolongo}}\ and\ \bibinfo {author} {\bibfnamefont {N}~\bibnamefont
  {Marzari}},\ }\bibfield  {title} {\enquote {\bibinfo {title} {Ionic
  correlations and failure of nernst-einstein relation in solid-state
  electrolytes},}\ }\href {\doibase 10.1103/physrevmaterials.1.025402}
  {\bibfield  {journal} {\bibinfo  {journal} {Phys. Rev. Mater.}\ }\textbf
  {\bibinfo {volume} {1}} (\bibinfo {year} {2017}),\
  10.1103/physrevmaterials.1.025402}\BibitemShut {NoStop}%
\bibitem [{\citenamefont {Kahle}\ \emph {et~al.}(2018)\citenamefont {Kahle},
  \citenamefont {Marcolongo},\ and\ \citenamefont {Marzari}}]{Kahle2018}%
  \BibitemOpen
  \bibfield  {author} {\bibinfo {author} {\bibfnamefont {L}~\bibnamefont
  {Kahle}}, \bibinfo {author} {\bibfnamefont {A}~\bibnamefont {Marcolongo}}, \
  and\ \bibinfo {author} {\bibfnamefont {N}~\bibnamefont {Marzari}},\
  }\bibfield  {title} {\enquote {\bibinfo {title} {Modeling lithium-ion
  solid-state electrolytes with a pinball model},}\ }\href {\doibase
  10.1103/physrevmaterials.2.065405} {\bibfield  {journal} {\bibinfo  {journal}
  {Phys. Rev. Mater.}\ }\textbf {\bibinfo {volume} {2}} (\bibinfo {year}
  {2018}),\ 10.1103/physrevmaterials.2.065405}\BibitemShut {NoStop}%
\bibitem [{\citenamefont {Resta}\ and\ \citenamefont
  {Vanderbilt}(2007)}]{Resta-Vanderbilt2007}%
  \BibitemOpen
  \bibfield  {author} {\bibinfo {author} {\bibfnamefont {R}~\bibnamefont
  {Resta}}\ and\ \bibinfo {author} {\bibfnamefont {D}~\bibnamefont
  {Vanderbilt}},\ }\enquote {\bibinfo {title} {Theory of polarization: A modern
  approach},}\ in\ \href {\doibase 10.1007/978-3-540-34591-6_2} {\emph
  {\bibinfo {booktitle} {Physics of Ferroelectrics: A Modern Perspective}}}\
  (\bibinfo  {publisher} {Springer Berlin Heidelberg},\ \bibinfo {address}
  {Berlin, Heidelberg},\ \bibinfo {year} {2007})\BibitemShut {NoStop}%
\end{thebibliography}

%

\end{document}